\documentclass[12pt]{article}
 
\usepackage[]{natbib}
\usepackage[]{aasms4}
\usepackage[]{graphicx}
 
\begin{document}
\title{Occultations of astrophysical radio sources as probes of planetary environments: A case study of Jupiter and possible applications to exoplanets}

\author{Paul Withers$^{a,b,*}$ and Marissa~F. Vogt$^{b}$} 
\affil{$^{a}$ Astronomy Department, Boston University, 725 Commonwealth Avenue, Boston, MA 02215, USA}
\affil{$^{b}$ Center for Space Physics, Boston University, 725 Commonwealth Avenue, Boston, MA 02215, USA}



\pagebreak



\pagebreak

\begin{abstract}
Properties of planetary atmospheres, ionospheres, and magnetospheres are difficult to measure from Earth.
Radio occultations are a common method for measuring these properties, 
but they traditionally rely on radio transmissions from a spacecraft near the planet.
Here we explore whether occultations of radio emissions from a distant astrophysical radio source 
can be used to measure magnetic field strength, plasma density, and neutral density around planets.
In a theoretical case study of Jupiter, we find that significant changes in polarization angle due to Faraday rotation occur for radio signals that pass within 10 Jupiter radii of the planet and that significant changes in frequency and power occur from radio signals that pass through the neutral atmosphere.
There are sufficient candidate radio sources, such as pulsars, active galactic nuclei, and masers, 
that occultations are likely to occur at least once per year.
For pulsars, time delays in the arrival of their emitted pulses can be used to measure plasma density.
Exoplanets, whose physical properties are very challenging to observe, may also occult distant astrophysical radio sources, such as their parent stars.
\end{abstract}

\noindent

\pagebreak

\section{\label{sec:intro}Introduction}

The passage of a radio signal through the atmosphere, ionosphere, or magnetosphere of a planet 
affects the polarization, frequency, and power of the radio signal
\citep{fjeldbo1965a, eshleman1973, budden1985}.
In planetary science, radio occultation experiments have used these effects to measure properties of planetary atmospheres, ionospheres, and magnetospheres
\citep{eshleman1973, tyler1987, howard1992, withers2010}.
A radio occultation experiment typically involves the transmission of a radio signal from a spacecraft located near a planet to Earth. 
Such experiments rely on the presence of a spacecraft near the planet to provide a source for the radio signal.
Here we explore whether similar measurements of planetary environments can be made using distant astrophysical radio sources.
We use ``planetary environment'' to encompass the atmosphere, ionosphere, and magnetosphere of a planet.
For conceptual simplicity, we discuss effects on the radio signal in terms of geometrical optics.
Wave optics methods exist for interpreting radio occultation observations \citep{karayel1997}.

Occultations of distant astrophysical sources have a successful history in solar system studies. 
At radio wavelengths, occultations of pulsars by the Sun have measured properties of the solar corona 
\citep{counselman1972, counselman1973, weisberg1976, bird1980, ord2007}.
At visible and infra-red wavelengths, occultations of stars by planets and their satellites have measured properties of their atmospheres and rings
\citep{elliot1977, bouchez2003, sicardy2006, sicardy2016, diasoliveira2015}.

Sections~\ref{sec:pol}--\ref{sec:power} summarize how the polarization angle, frequency, and power of a radio signal
 are affected by conditions in the planetary environment.
Section~\ref{sec:jupiter} describes the predicted effects on a radio signal that propagates through the environment of Jupiter.
Section~\ref{sec:sources} discusses suitable distant astrophysical radio sources.
Section~\ref{sec:pulsars} considers effects on the time delay of pulses emitted by a pulsar.
Section~\ref{sec:occrate} evaluates the likelihood of Jupiter occulting a suitable distant astrophysical radio source.
Section~\ref{sec:exo} addresses potential applications to exoplanets and stars.
Section~\ref{sec:concs} presents the conclusions of this work.

\section{\label{sec:pol}Polarization Angle}

The plane of polarization of a radio signal is sensitive to the plasma density and magnetic field along the ray path.
Faraday rotation alters the polarization angle of a radio signal that passes through a plasma in the presence of a magnetic field.
The angle of rotation of the plane of polarization, $\phi$ \emph{(radians)}, satisfies
\citep{born1959, jackson1975}:

\begin{eqnarray}
\label{eqn:faraday01}
\phi = \frac{e^{3}}{8 \pi^{2} m_{e}^{2} \epsilon_{0} c f^{2}} \int N \underline{B} \cdot \underline{dl}
\end{eqnarray}

where $-e$ is the electron charge, $m_{e}$ is the electron mass, $\epsilon_{0}$ is the permittivity of free space, $c$ is the speed of light, 
$f$ is the frequency of the radio signal, $N$ is the electron density, $\underline{B}$ is the magnetic field, the integral is taken along the path of the radio signal, and $f$ is much greater than the gyrofrequency or plasma frequency \citep{howard1992}.
Equation~\ref{eqn:faraday01} can be used to relate the plane of polarization of a radio signal to the plasma density and magnetic field along the ray path.


\section{\label{sec:freq}Frequency}

A radio signal that passes through a planetary environment is refracted by plasma and neutrals along its path.
This affects the received frequency of the radio signal.
The refractive index of a planetary environment, $\mu$, can be expressed in terms of the refractivity, $\nu$, which is defined as $\mu = 1 + \nu$
\citep{withers2010}.
In a neutral atmosphere, $\nu$ is proportional to the neutral number density, $n$, and satisfies $\nu_{n} = \kappa n$, where $\kappa$ is the refractive volume of the atmospheric constituents
\citep{withers2010}. For many typical constituents, $\kappa$ is on the order of $10^{-29}$ m$^{-3}$
\citep{withers2010}.
In an ionosphere or magnetosphere, $\nu$ is proportional to the electron density, $N$, and satisfies
\citep{budden1985, withers2010}:

\begin{eqnarray}
\label{eqn:nuforplasma}
\nu_{e} = - \frac{N e^2}{8 \pi^{2} m_{e} \epsilon_{0} f^{2}}
\end{eqnarray}

In practice, refractivity is generally contributed by either neutrals or plasma, not both, as regions of high plasma density and high neutral density are well separated \citep{withers2010}.
Refraction causes the path of the radio signal to bend. This bending is described by the bending angle, $\alpha$, which is the angle between the asymptotes of the ray path entering and exiting the planetary environment.
If the planetary environment is spherically symmetric, then the bending angle $\alpha$ \emph{(radians)} satisfies
\citep{fjeldbo1971}:

\begin{eqnarray}
\label{eqn:adeqn02}
\alpha_{j} \left(a_{j}\right) = - 2 a_{j} \int_{r=r_{j}}^{r=\infty} 
\frac{d \ln \mu\left(r\right)}{dr} \frac{dr}{\sqrt{ \left(\mu\left(r\right) r\right)^{2} - a_{j}^{2}}}
\end{eqnarray}

where $r$ is radius from the center of the planet and subscripts indicate a specific ray path, which is usually equivalent to a specific time at the receiver.
Here we have adopted the convention of \citet{fjeldbo1971} and \citet{withers2014} that positive bending is away from the planet such that refraction in the neutral atmosphere leads to a negative bending angle $\alpha$. \citet{ahmad1998} and \citet{withers2010} adopted the opposite convention.
The impact parameter $a$ and closest approach distance $r$ are related by Bouguer's rule, $\mu r = a$
\citep{born1959, fjeldbo1971, eshleman1973, withers2010}.
If $\nu$ varies exponentially with radial distance and has scale height $H$, then $\alpha$ satisfies \citep{eshleman1979, yakovlev2002, withers2010}:

\begin{eqnarray}
\label{eqn:expalpha}
\alpha = -\nu \sqrt{2 \pi a / H}
\end{eqnarray}

The minus sign in this equation is not present in the corresponding equation in \citet{withers2010} (their Equation 35) due to the sign convention for $\alpha$ noted above.
Since the refracted ray path differs from what it would be in a vacuum, the rate of change of the length of the refracted ray path also differs from what it would be in a vacuum.
This affects the Doppler shift experienced by the radio signal.
The frequency of the received radio signal differs from its vacuum value by the frequency residual, $\Delta f$, which satisfies
\citep{hinson1999, withers2014}:

\begin{eqnarray}
\label{eqn:dfeqn01}
\left|\frac{\Delta f}{f}\right| \approx \left|\frac{V_{\perp} \alpha}{c}\right|
\end{eqnarray}

where 
$V_{\perp}$ is the component of the velocity of the transmitter relative to the receiver that is perpendicular to the ray path.
This equation is commonly stated without emphasizing the sign of $\Delta f$.
A signed version of this equation is given in the Appendix of \citet{withers2014} (their Equation 30). 
This shows that the frequency residual caused by refraction in the neutral atmosphere during ingress into an occultation is negative, whilst the corresponding egress residual is positive. This is illustrated in Figure 2 of \citet{lindal1983} for the Voyager 1 Titan occultation.
Equations~\ref{eqn:nuforplasma}--\ref{eqn:dfeqn01} can be used to relate the frequency of a radio signal to the plasma density or neutral density along the ray path.

\section{\label{sec:power}Power}


The power of a radio signal is sensitive to the plasma density and neutral density along the ray path.
Several factors associated with the propagation of the radio signal through a planetary environment can reduce the received power.
These include refractive defocusing, absorption by neutral species, and absorption by plasma via electron-neutral collisions
\citep{eshleman1973, howard1992, kliore2004, withers2011}.
Here we concentrate on the strongest of these factors, defocusing.
In a neutral atmosphere, changes in the bending angle $\alpha$ with impact parameter $a$ causes the beam to spread out, which reduces its intensity.
We define $I$ as the ratio of the received intensity to the intensity that would have been received in the absence of refraction during the propagation of the radio signal through a planetary environment. $I$ satisfies \citep{eshleman1980}:

\begin{eqnarray}
\label{eqn:defocusingloss01}
I^{-1} = 1 + D \frac{d \alpha}{da}
\end{eqnarray}

Here $D$ is the distance from the planet to the closest of the receiver and transmitter.
Note that the corresponding equation in \citep{eshleman1980} (their Equation A6) contains a minus sign in place of 
Equation~\ref{eqn:defocusingloss01}'s plus sign due to differences in sign convention for $\alpha$.
In a neutral atmosphere, bending angle $\alpha$ becomes increasingly negative as impact parameter $a$ decreases, which ensures that $I^{-1}$ is greater than 1.
That is, $\alpha$ is negative and its magnitude increases deeper in the atmosphere.
Thus $I$ is positive and less than 1, indicating a reduction in intensity.
For a spacecraft radio occultation, $D$ is the distance from the planet to the nearby spacecraft.
For an occultation of a distant astrophysical radio source, this is the distance from the planet to Earth, which is much larger.
Thus defocusing losses will be significantly greater for occultation of a distant astrophysical radio source
than for a spacecraft radio occultation.

\emph{The large distance $D$ also affects the spatial resolution of the observations.
In the geometrical optics approach used here, the vertical resolution of the observations is diffraction-limited to the size of the Fresnel zone, 
$2 \sqrt{\lambda D}$, where $\lambda$ is wavelength \citep{karayel1997}.
For spacecraft occultations of solar system objects, this is typically on the order of 1 km.
For occultations of a distant astrophysical radio source, the vertical resolution will be appreciably larger. 
Note that existing wave optics methods of analysis can overcome the diffraction limit \citep{karayel1997}.}

In a spacecraft radio occultation, attenuation due to propagation through the magnetosphere and ionosphere are neglible as $\left|d \alpha/da\right|$ are small. In an occultation of a distant astrophysical radio source, $\left|d \alpha/da\right|$ in the magnetosphere and ionosphere remains small, but the increased size of $D$ means that refractive effects on the intensity $I$ are not necessarily neglible.
Moreover, in a magnetosphere or ionosphere, $\alpha$ may decrease as $a$ increases.
In this case, $I$ is not necessarily positive and less than 1.
$I$ may be positive and greater than 1, indicating an increase in intensity (focusing).
Or $I$ may be negative, which indicates the presence of multipath effects in which multiple rays reach the receiver simultaneously.
Equation~\ref{eqn:defocusingloss01} can be used to relate the power of a radio signal to the plasma density or neutral density along the ray path.


\section{\label{sec:jupiter}Simulated effects at Jupiter}






We consider Jupiter as a test case to assess the significance of predicted changes in polarization angle, frequency, and power of radio emissions from a distant astrophysical radio source during an occultation of the radio source by a planet.
A distinct advantage of occultation-based techniques is that effects caused by propagation through the interplanetary or interstellar medium can be eliminated by using pre- and post-occultation measurements as a baseline. 
Consequently, we do not need to consider such effects here.
\emph{However, effects caused by propagation through the terrestrial atmosphere and ionosphere vary on timescales short by comparison to the duration of an occultation. These effects must be removed by independent simultaneous observations of the terrestrial environment, which are routinely acquired at the NASA Deep Space Network stations and other radio observatories. Validated empirical models of the terrestrial environment are also available.}

For simplicity, we initially assume that all aspects of the planetary environment, including the atmosphere, ionosphere, and magnetosphere, are spherically symmetric and adopt an illustrative model of Jupiter's environment.
We assume that the neutral density in the atmosphere $n\left(r\right) = n_{0} \exp -\left(r - 1 R_{J}\right)/H$, where $n_{0} = 4 \times 10^{25}$ m$^{-3}$, $H$ = 35 km, and 1 $R_{J}$ is the radius of Jupiter to the 1 bar pressure level ($7 \times 10^{4}$ km)
\citep{seiff1998}.
Jupiter's neutral atmosphere is oblate, not spherical
\citep{hubbard1975, kliore1975b, kliore1976}, but the assumption of spherical symmetry is reasonable for this exploratory work.
The assumption of an isothermal atmosphere is also only weakly valid \citep{seiff1998}.
The dependence of the assumed electron density in the ionosphere and magnetosphere $N\left(r\right)$ on radial distance is more complicated.
In the ionosphere, we assume that $N\left(r\right)$ is a Chapman function, $N = N_{0} \exp \left(1 - x - \exp\left(-x\right)\right)$ 
where $N_{0} = 10^{11}$  m$^{-3}$, $x = \left(r - r_{p}\right) / H_{p}$, $r_{p}$ = 1750 km above 1 $R_{J}$, and $H_{p}$ = 350 km
\citep{yelle2004}.
The assumption of spherical symmetry is not realistic for the ionosphere, where densities vary with solar zenith angle.
The assumed ionospheric structure is based on observations close to the terminator.
We assume that the top of the ionosphere occurs where the electron density equals $10^{9}$ m$^{-3}$, which is an altitude of 5300 km above 1 $R_{J}$ ($r = 1.07 R_{J}$).
Since there are few observations of electron density between the top of the ionosphere and the inner boundary of the magnetosphere at 10 $R_{J}$, we adopt a simple bridging function in which $N$ decreases exponentially with $r$ such that $N=10^{8}$ m$^{-3}$ at 10 $R_{J}$. This corresponds to a scale height of $2.7 \times 10^{5}$ km or about 4 $R_{J}$.
In the inner magnetosphere, which begins at 10 $R_{J}$, 
$N$ decreases exponentially with $r$ with a scale height of 2.5 $R_{J}$ \citep{bagenal2011}.
In the outer magnetosphere, $N$ decreases exponentially with $r$ with a scale height of 40 $R_{J}$ \citep{bagenal2011}.
The outer boundary of the magnetosphere occurs at 100 $R_{J}$ and an electron density of $10^{4}$ m$^{-3}$ \citep{bagenal2011}. 
The transition from the inner magnetosphere to the outer magnetosphere occurs where their respective electron density functions intersect, namely $6 \times 10^{4}$ m$^{-3}$ and 28.6 $R_{J}$.
The assumption of spherical symmetry is not realistic for the magnetosphere. 
Plasma in Jupiter's middle and outer magnetosphere is largely confined to a plasma sheet of thickness $\sim$5--10 $R_{J}$ that lies near the equatorial plane \citep{behannon1981, khurana2005}.
Furthermore, the size of Jupiter's magnetosphere is variable. The sunward distance to the outer boundary of the magnetosphere ranges from $\sim$60 to $\sim$90 $R_{J}$ and the terminator distance ranges from $\sim$90 to $\sim$130 $R_{J}$, depending on ambient solar wind conditions 
\citep{joy2002}. The tailward extent is discussed further in Section~\ref{sec:pulsars}.
We assume that the component of the magnetic field parallel to the line of sight, $\underline{B} \cdot \underline{dl} / dl$, which influences Faraday rotation of the plane of polarization, \emph{is independent of occultation geometry and equivalent to the magnitude of the magnetic field, $\left|B\right|$.}

We assume that the magnetic field decreases exponentially from $5 \times 10^{-4}$ T at 1 $R_{J}$ to $5 \times 10^{-8}$ T at 20 $R_{J}$
\citep{acuna1983}, then decreases exponentially from $5 \times 10^{-8}$ T at 20 $R_{J}$ to $5 \times 10^{-9}$ T at 100 $R_{J}$ \citep{kivelson2002}. 
The magnetic field strength at radial distances less than 20 $R_{J}$ is dominated by the planet's internal magnetic field.
The magnetic field strength at larger distances is significantly influenced by magnetospheric currents.
The assumptions of spherically symmetric magnetic field strength and $\underline{B} \cdot \underline{dl} = \left|B\right| dl$ are convenient in that they ensure that the resultant Faraday rotation depends only on the radial distance of closest approach, not any additional geographical considerations such as the latitude of closest approach or the direction of the ray path. However, these assumptions are not particularly well satisfied for Jupiter's predominantly dipolar field. 

\begin{figure}
\noindent\includegraphics[width=15cm,keepaspectratio]{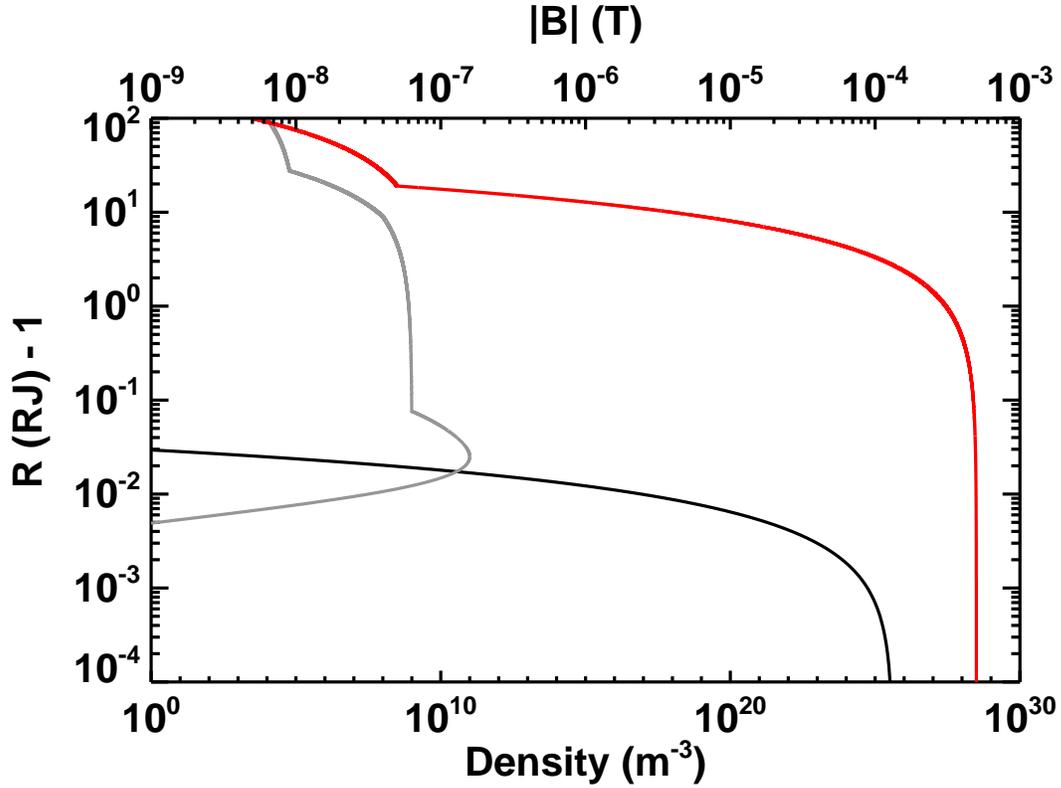}
\caption{\label{fig:allinputs}
Simulated neutral density (black), electron density (grey), and magnetic field strength (red) in the environment of Jupiter.
Densities are indicated on the lower axis and magnetic field strength is indicated on the upper axis.
Discontinuities in the radial gradient in plasma density occur at $r = 1.07 R_{J}, 10 R_{J}$, and $28.6 R_{J}$.
A discontinuity in the radial gradient in magnetic field strength occurs at $r = 20 R_{J}$.
}
\end{figure}

The modeled densities and field strength are shown in Figure~\ref{fig:allinputs}.
Radial distances between 1 $R_{J}$ and 100 $R_{J}$ are shown, where $R_{J}$ is the radius of Jupiter to the 1 bar pressure level ($7 \times 10^{4}$ km).
Note that radio signals can propagate deeper into the planet than the 1 bar pressure level (1 $R_{J}$), although these results are not shown.
This model includes some gross simplifications, but it offers a reasonable representation of conditions near the terminator
of Jupiter given the assumption of spherical symmetry. That is sufficient for the exploratory purposes of this work.

We assume 
that $V_{\perp}$ = 17 km s$^{-1}$ to find the frequency shift (Equation~\ref{eqn:dfeqn01}) and
that $D$ = 5 AU to find the power loss (Equation~\ref{eqn:defocusingloss01}).
This speed $V_{\perp}$ is the difference between Earth's 30 km s$^{-1}$ orbital speed and Jupiter's 13 km s$^{-1}$ orbital speed, so we implicitly assume that the distant astrophysical radio source is at rest. 
\emph{Smaller speeds and longer occultation durations are possible if Jupiter is at opposition and in retrograde motion.}
With the assumed speed of 17 km s$^{-1}$, the impact parameter $a$ changes by $\sim$1 $R_{J}$ per hour. The impact parameter $a$ will pass through ionospheric and atmospheric altitudes on a timescale of minutes, which is amenable to observations.
Yet $a$ will pass through inner magnetospheric distances on timescales of tens of hours. 
Due to Earth's rotation, this timescale would preclude observations by a single ground-based observer and would require coordination across multiple observing sites.
Furthermore, since Jupiter has a ten hour rotational period, the magnetospheric parts of an occultation will span multiple Jupiter rotations.
\emph{This has the potential to introduce significant complexity to the analysis of observations, as the assumption of spherical symmetry is flawed for the magnetosphere.}

Propagation through a planetary environment has frequency-dependent effects
on the polarization angle, frequency, and power of a radio signal.
We therefore consider multiple radio frequencies in this case study --- 0.1, 1, and 10 GHz.
Radio astronomers routinely observe astrophysical sources in this range of frequencies.

\emph{For our chosen $D$ and $f$, the Fresnel zone size is 3000 km (0.1 GHz), 1000 km (1 GHz), and 300 km (10 GHz) (Section~\ref{sec:power}).
This is the diffraction-limited vertical resolution that can be achieved using geometrical optics methods.
Existing wave optics methods of analysis can be used to improve the vertical resolution by an order of magnitude, thereby yielding sub-scale height resolution \citep{karayel1997}.}

\begin{figure}
\noindent\includegraphics[width=15cm,keepaspectratio]{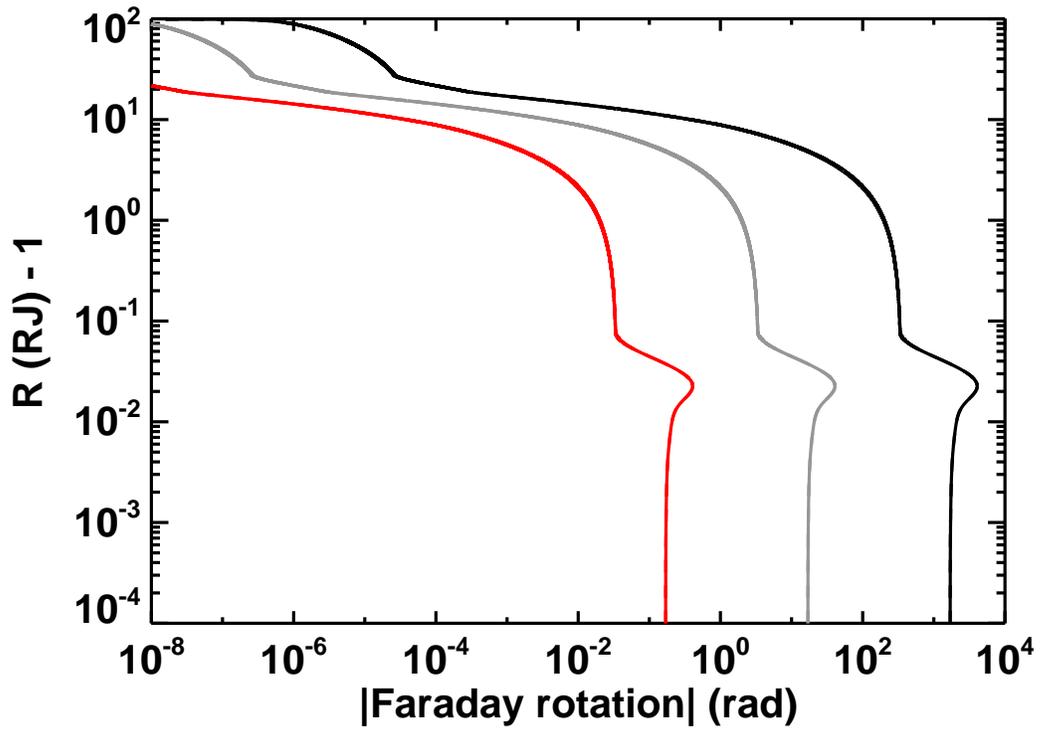}
\caption{\label{fig:intfaradayvslogr}
Simulated changes in the polarization angle of a radio signal as a function of closest approach distance.
Frequencies of 0.1, 1, and 10 GHz are shown in black, grey, and red, respectively.
}
\end{figure}

Polarization angle results are shown in Figure~\ref{fig:intfaradayvslogr}.
Faraday rotation is inversely proportional to the square of frequency (Equation~\ref{eqn:faraday01}).
Polarization angle changes in excess of 1 rad are present for 0.1 GHz (1 GHz) radio signals that pass within 10 $R_{J}$ (3 $R_{J}$) of the center of the planet.
These are caused by magnetospheric plasma. Effects are larger for radio signals that pass through the ionosphere. They exceed 0.1 rad even at 10 GHz.
These results use the idealized magnetic conditions described above. As previously stated, a more realistic model would decrease these idealized 
polarization angle changes by a factor of 0.2.
Since the polarization angle can only be determined within the range 0--$\pi$ radians, note that changes in excess of $\pi$ radians must be tracked by observing at high time resolution.

\begin{figure}
\noindent\includegraphics[width=15cm,keepaspectratio]{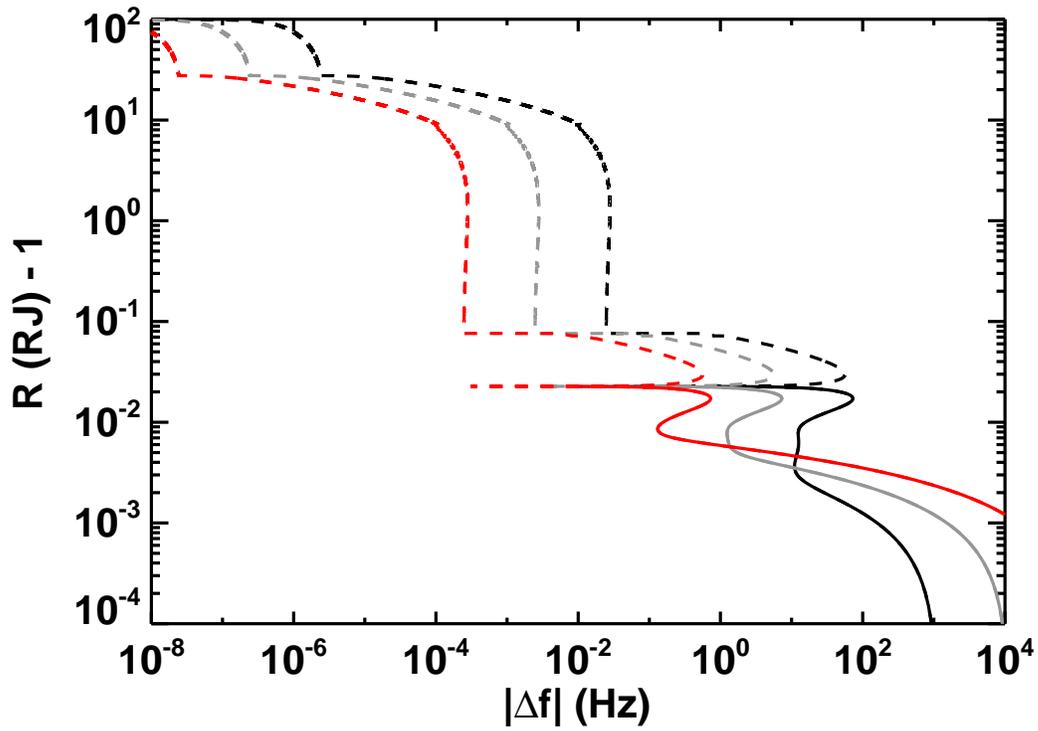}
\caption{\label{fig:deltafvslogr}
Magnitude of simulated frequency residual $\Delta f$ as a function of closest approach distance \emph{for an egress occultation.}
Frequencies of 0.1, 1, and 10 GHz are shown in black, grey, and red, respectively.
Negative and positive frequency shifts are shown as dashed and solid lines, respectively.
}
\end{figure}

Frequency shift results are shown in Figure~\ref{fig:deltafvslogr}.
As discussed in Section~\ref{sec:freq}, the sign of the frequency residual changes between ingress and egress: egress is shown here.
Furthermore, this assumes a deep occultation in which the line of sight between the distant astrophysical radio source and Earth is completely occulted by the planet at the center of the occultation. 
Behavior around the center of a ``grazing'' occultation, such as an occultation in which the closest approach distance never drops below 2 $R_{J}$,
is more complex, as is clear from the requirement that the sign of the frequency residual change from ingress to egress. 
The simplest explanation is that the velocity component $V_{\perp}$ in Equation~\ref{eqn:dfeqn01}, which is the proportionality relationship between the frequency residual and the bending angle, is zero at the center of the occultation.
Further details are available in Appendix A of \citet{withers2014}.

Returning to Figure~\ref{fig:deltafvslogr} and the case of a deep occultation,
magnetospheric plasma causes the frequency shift in the 0.1 GHz radio signal to exceed $10^{-2}$ Hz (1 part in $10^{10}$) for distances less than 10 $R_{J}$.
At the ionospheric peak, the frequency shift in the 0.1 GHz radio signal is on the order of 100 Hz (1 part in $10^{6}$). 
The neutral atmosphere has the greatest effect on the frequency. 
The frequency shift exceeds 1 kHz for all three radio frequencies before the 1 bar level is reached.

\begin{figure}
\noindent\includegraphics[width=15cm,keepaspectratio]{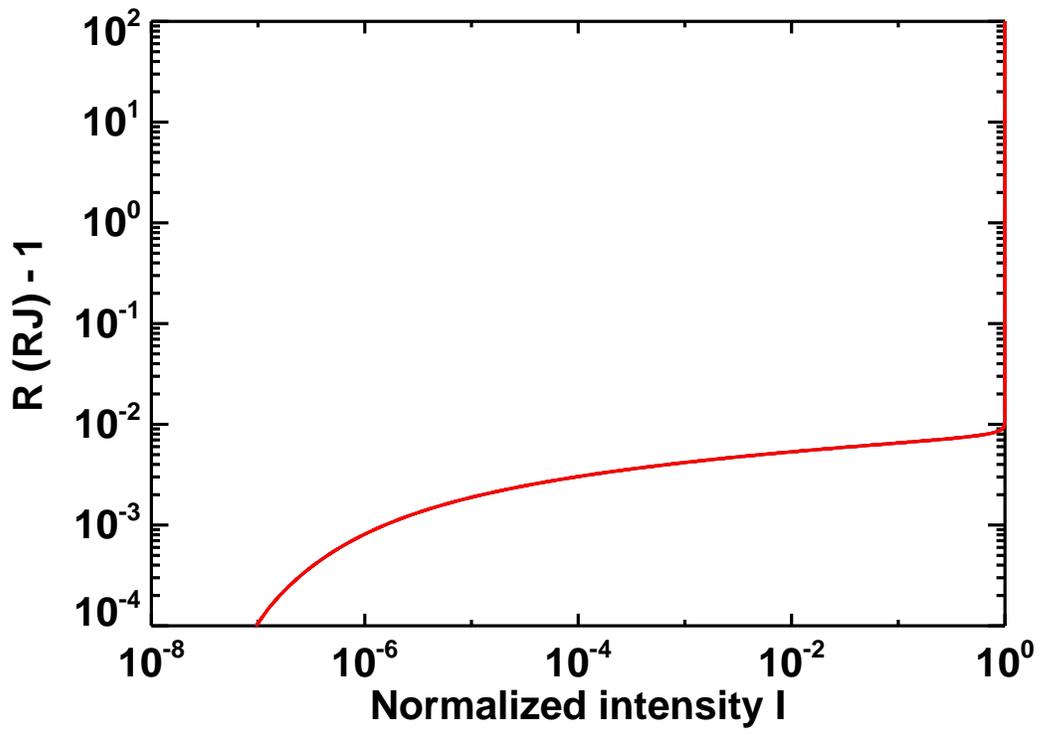}
\caption{\label{fig:powerdbvslogr}
Frequency-independent normalized intensity $I$ as a function of closest approach distance considering only the neutral atmosphere.
}
\end{figure}

Refractive effects on the power are negligible for rays whose closest approach distance is in the magnetosphere.
Refractive effects on the power are complicated for rays whose closest approach distance is in the ionosphere.
Individual rays can be focused, which increases their intensity, and multipath effects can also occur.
Consequently, Figure~\ref{fig:powerdbvslogr} shows power losses calculated assuming refraction in the neutral atmosphere only.
With this assumption, Equations~\ref{eqn:expalpha} and \ref{eqn:defocusingloss01} lead to:

\begin{eqnarray}
\label{eqn:expdefocusing01}
I^{-1} = 1 - \frac{D \alpha}{H}
\end{eqnarray}

Since the observable quantity $I^{-1} - 1$ is proportional to $\alpha$, which is proportional to $\exp \left(-a/H\right)$ 
(Equation~\ref{eqn:expalpha}), the neutral scale height $H$ can be inferred from measurements of $I\left(a\right)$.
However, these effects from the neutral atmosphere only occur at altitudes less than 700 km above 1 $R_{J}$.
This requires the occulted radio source to have a very close approach to Jupiter.

If data analogous to those shown in Figures~\ref{fig:intfaradayvslogr}--\ref{fig:powerdbvslogr} are acquired during an occultation of 
a distant astrophysical radio source by a planet, then they can be analyzed using the equations set forth above to determine how planetary magnetic field strength, electron density, and neutral density vary with radial distance.


\section{\label{sec:sources}Suitable distant astrophysical radio sources}

The ideal distant astrophysical radio source for occultation measurements of a planetary environment is point-like.
Furthermore, its emissions should have a large and constant polarization, a narrow and constant frequency, and a large and constant power.
Nevertheless, a distant astrophysical radio source that does not satisfy all these requirements may still be useful. 
For instance, changes in frequency and power during an occultation will be informative even if emissions from distant astrophysical radio source are unpolarized.
The angular size of the distant astrophysical radio source should be no larger than the angular size over which the environment of the occulted planet is changing.
For the Jupiter case study, that corresponds to an angular size of 
0.01 arcseconds for the atmosphere, 0.1 arcseconds for the ionosphere, 
and 
50 arcseconds for the magnetosphere.
Candidate distant astrophysical radio sources include pulsars, active galactic nuclei, and masers.

Pulsars are neutron stars with diameters on the order of 10 km that can be considered to be point-like sources \citep{burke2010}.
Their radio emissions are \emph{often} strongly polarized \citep{burke2010} and
consist of periodic short pulses \citep{carroll2007}.
Although the power in an individual pulse is highly variable, the power in an integrated pulse profile, which is the average of $\sim$100 pulses, is quite stable \citep{carroll2007}.
Since pulsar periods are on the order of milliseconds to seconds \citep{carroll2007}, it is feasible to generate an integrated pulse profile for which each constituent pulse can be considered to have propagated along the same path through the planetary ionosphere and magnetosphere.

Active galactic nuclei are supermassive black holes at the centers of galaxies \citep{carroll2007}.
Accretion onto the black hole generates intense emissions at radio wavelengths \citep{carroll2007}.
The compact central source region has an angular size on the order of milliarcseconds \citep{carroll2007}, so can be considered as a point-like size for occultations of Jupiter's environment. Emissions from the central source can be strongly linearly polarized \citep{carroll2007}.
High resolution imaging is required to exclude emissions from lobes and jets emanating from the central source \citep{carroll2007}.
Emissions from some classes of active galactic nuclei include narrow spectral lines \citep{carroll2007}.

Masers can be either galactic or extra-galactic \citep{gray1999, burke2010}.
Galactic masers are most commonly found in association with stars \citep{gray1999, burke2010}.
Since the spatial extent of a region of maser emission is on the order of a few AU \citep{gray1999, burke2010}, 
the angular size of a galactic maser can be considered point-like for occultations of Jupiter's environment.
Their emissions are often strongly polarized \citep{gray1999, burke2010}.
Galactic maser emission occurs in narrow lines that correspond to energy level transitions in species such as SiO, H$_{2}$O, and OH \citep{gray1999, burke2010}.
Extra-galactic masers, known as megamasers, are associated with active galactic nuclei \citep{lo2005}.
They are less polarized than galactic masers \citep{gray1999}.
Their emissions, which are relatively broad \citep{gray1999}, come from OH or H$_{2}$O lines \citep{lo2005}.




\section{\label{sec:pulsars}Time delay of pulsar emissions}

Another method for measuring the local electron density at a planet is available for occultations of pulsars. 
The arrival time for a given pulse is delayed relative to the vacuum arrival time due to the pulse's propagation through plasma in the interstellar medium and planetary environment.
Thus pulses that pass through regions of high electron density in a planetary magnetosphere and ionosphere will be received later than expected.
The frequency-dependent time delay, $T$, satisfies
\citep{croft1971, budden1985, howard1992}:

\begin{eqnarray}
\label{eqn:timedelay01}
T = \frac{e^{2}}{8 \pi^{2} m \epsilon_{0} c f^{2}} \int N dl
\end{eqnarray}

\begin{figure}
\noindent\includegraphics[width=15cm,keepaspectratio]{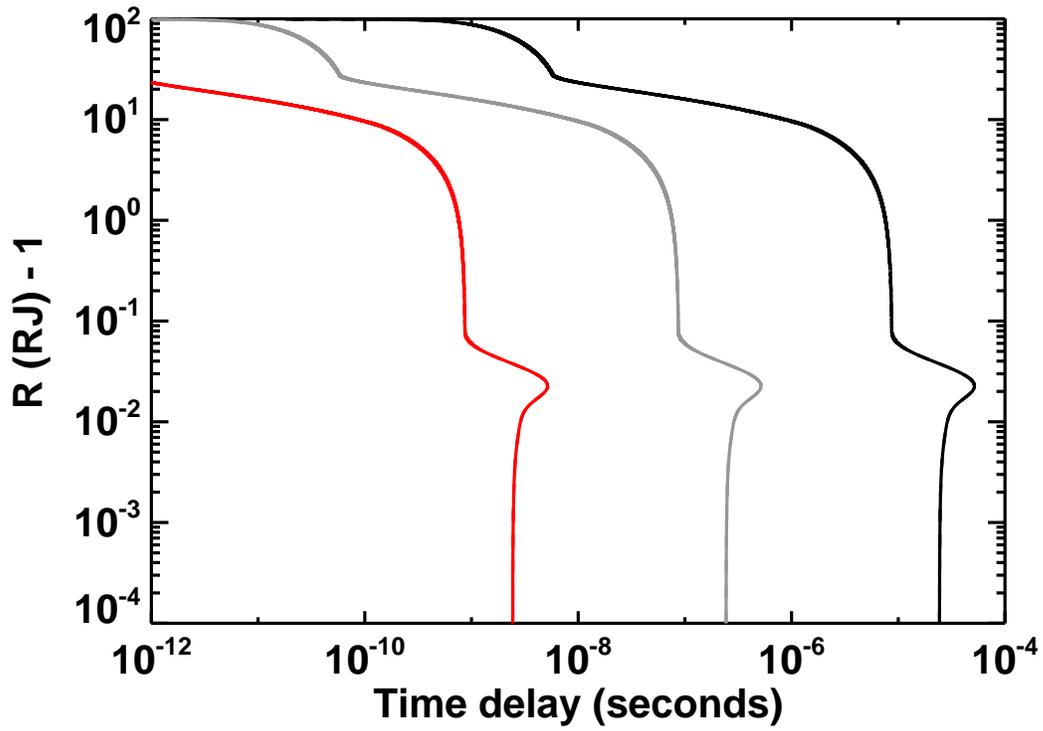}
\caption{\label{fig:timedelayvslogr}
Simulated time delay as a function of closest approach distance for occultation that is not at opposition.
Frequencies of 0.1, 1, and 10 GHz are shown in black, grey, and red, respectively.
}
\end{figure}

Simulated time delays relative to vacuum for the spherically-symmetric Jupiter case study are shown in Figure~\ref{fig:timedelayvslogr}.
The time delay exceeds 1 \emph{microsecond} for 0.1 GHz radio signals for distances less than 10 $R_{J}$.
Averaging over a large number of pulses will increase the precision with which the time delay can be measured.
A time series of time delays provides a series of column-integrated electron densities (m$^{-2}$) for different distances of closest approach.
These can be inverted to provide a radial profile of the local electron density (m$^{-3}$) \citep{quemerais2006}.

The spherically symmetric model of Jupiter's environment used in Figure~\ref{fig:timedelayvslogr} is appropriate for conditions 
near the terminator and on the sunward side of the planet.
On the anti-sunward side of the planet, 
Jupiter's magnetosphere is so elongated that Saturn's orbit, which is twice as far away from the Sun as Jupiter's, is regularly within Jupiter's magnetotail \citep{lepping1983}.
On the anti-sunward side, 
Jupiter's magnetosphere can be represented as a cylinder of length 15 AU ($3 \times 10^{4} R_{J}$) \citep{lepping1983} 
and radius 200 $R_{J}$  \citep{khurana2004}.
In this representation, Jupiter is at the center of one circular face and the cylinder extends outwards away from the Sun.
The electron density in the elongated magnetotail is $5 \times 10^{3}$ m$^{-3}$ \citep{nicolaou2015}, 
20 times smaller than the $10^{5}$ m$^{-3}$ density in the surrounding solar wind \citep{ebert2014}. 


Consequently, the simulated time delays shown in Figure~\ref{fig:timedelayvslogr} are inaccurate for the case in which the radio signal has traveled along the Sun-Jupiter line, as would occur for an occultation at opposition.
For this case, we model plasma densities on the sunward side of Jupiter as before, but model plasma densities on the anti-sunward side of Jupiter with the elongated magnetotail described above.
For closest approach distances that on the sunward side are in the outer magnetosphere, the net effect is a negative time delay, or a time advance.
The long, low density magnetotail causes a time advance relative to propagation on a parallel path through the adjacent solar wind plasma, and higher densities encountered on the short sunward side are not high enough to compensate.
By contrast, for closest approach distances that are close to the planet, the net effect is a time delay due to the extremely high densities encountered on the short sunward side.

\begin{figure}
\noindent\includegraphics[width=15cm,keepaspectratio]{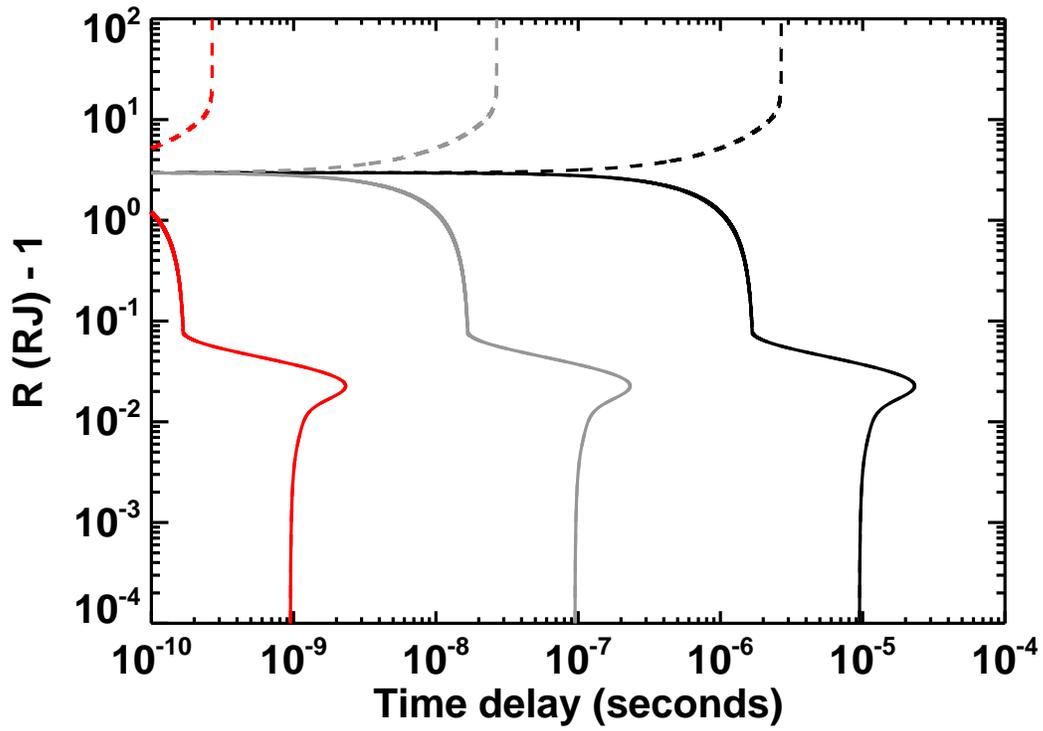}
\caption{\label{fig:timedelayoppovslogr}
Simulated time delay as a function of closest approach distance for occultation at opposition.
Negative time delays (i.e., time advances) are shown as dashed lines.
Frequencies of 0.1, 1, and 10 GHz are shown in black, grey, and red, respectively.
}
\end{figure}

Simulated time delays for this case are shown in Figure~\ref{fig:timedelayoppovslogr}.
These time delays are relative to a parallel path through the solar wind plasma that surrounds Jupiter's environment.
The magnitude of the time delay exceeds 1 \emph{microsecond} for 0.1 GHz radio signals for distances less than 2 $R_{J}$ or greater than 6 $R_{J}$.
As for the non-opposition case, a time series of time delays for different distances of closest approach, as is illustrated in 
Figure~\ref{fig:timedelayoppovslogr}, provides information about the spatial distribution of plasma density around the planet.
Note also that effects exist at $R = 100$--$200 R_{J}$ that are not shown on this figure.
At these distances of closest approach, the time advance is the same as for tens of $R_{J}$ --- 3 \emph{microseconds}.
In this model, this 3 ms time advance ceases abruptly at 200 $R_{J}$ due to the sharp boundary between the low density magnetotail and the higher density solar wind at this distance from the central axis of the magnetotail. In reality, the boundary will have an appreciable spatial extent and the time advance will return to zero more gradually.
We do not consider effects of the magnetotail for the non-opposition case (Figure~\ref{fig:timedelayvslogr}) because the path length through the magnetotail is much reduced away from opposition.

\section{\label{sec:occrate}Likelihood of occurrence of Jupiter occultation}

In order to estimate the likelihood of Jupiter occulting a suitable distant astrophysical radio source, we estimate the fraction of the sky obscured from view by Jupiter as seen from Earth over the course of one orbital period of Jupiter (12 years).
Jupiter sweeps out a band of height $2 R_{J}$ and length $2 \pi a_{J}$ within a full sky area of $4 \pi a_{J}^{2}$, 
where $a_{J}$ is Jupiter's orbital semi-major axis.
Hence Jupiter occults $R_{J}/a_{J}$ or $10^{-4}$ of the sky over the course of one orbit.
Moreover, Figures~\ref{fig:intfaradayvslogr}--\ref{fig:deltafvslogr} shows that significant effects on polarization angle and frequency occur for radio signals with a closest approach distance of 10 $R_{J}$. This increases the fraction of sky covered to $10^{-3}$.
Since thousands of pulsars \citep{manchester2005}
and tens of thousands of active galactic nuclei \citep{krolik1999} have been identified,
it is likely that at least one occultation of a pulsar or an active galactic nucleus by the environment of Jupiter will occur over the course of one Jupiter year.
\emph{However, candidate distant astrophysical radio sources are concentrated in the galactic plane, not distributed isotropically, so occultations are most likely when Jupiter is near the intersection of the ecliptic and galactic planes.}


Occultations of Jupiter's magnetotail at opposition are also likely.
Since the projected area of a disk of radius 200 $R_{J}$ at a distance of 5 AU is $10^{-4}$ of the full sky, 
there is a reasonable prospect of a pulsar or other suitable distant astrophysical radio source occulting Jupiter's magnetotail at a given opposition.
Oppositions occur approximately once per Earth year.
A different region of the sky will be sampled at each opposition, which increases the likelihood that a suitable distant astrophysical radio source will be appropriately positioned for at least one opposition in the near future. 
\emph{The most promising oppositions are those that occur when Jupiter is near the concentration of distant astrophysical radio sources within the galactic plane.}


The Jupiter case study presents the problem of radio emissions from the planetary environment itself.
Jupiter emits radio noise at many frequencies, including decimetric radio noise at frequencies around 0.1--10 GHz \citep{bolton2004}
that must be distinguished from emissions from the distant astrophysical radio source.
These decimetric emissions are synchrotron radiation produced in the planet's intense radiation belts \citep{bolton2004}.
Their spectral irradiance is on the order of 10 Jy \citep{depater2003, bolton2004}, 
whereas a representative value for a pulsar is 10 mJy, three orders of magnitude smaller \citep{manchester2005}.

The source region for the decimetric emissions is approximately 1 $R_{J}$ in radius. At 5 AU, that corresponds to an angular extent of 20 arcseconds.
Imaging with an angular resolution of 20 milliarcseconds will reduce the observed intensity of the radiation belt emissions by 6 orders of magnitude, 
rendering them insignificant.
This issue is not a concern for other solar system planets.
Their decimetric emissions are much weaker than Jupiter's, since they lack radiation belts as intense as Jupiter's \citep{bolton2004}.

\section{\label{sec:exo}Application to exoplanets and stars}

The observations proposed in this article would be beneficial to studies of planets in our solar system.
Moreover, such measurements would be immensely valuable to studies of exoplanets beyond our solar system.
There are as yet no measurements of the magnetic field strength, electron density, or neutral atmospheric density in the vicinity of an exoplanet.
This technique could also detect ring systems around exoplanets if the rings are sufficiently dense to attenuate the radio signal appreciably 
\citep{kliore2004}.
This would be a radio analogue to the discovery of the rings of Uranus by their attenuation of starlight during 
a stellar occultation \citep{elliot1977}. 
Moons \emph{and atmosphere-less exoplanets}, if they fortuitously encounter the ray path, would also extinguish the radio signal.

The low probability of suitable alignment of Earth, an exoplanet, and a suitable distant astrophysical radio source presents a considerable challenge.
A Jupiter-like exoplanet in a Jupiter-like orbit that is seen face-on sweeps out an area of $10^{15}$ km$^{2}$ over the course of one orbit.
If that exoplanet is at a distance of 10 light-years, then it sweeps out only $10^{-14}$ of the total area of the sky.
If the exoplanet has a Jupiter-like magnetotail, then the magnetotail sweeps out $10^{-10}$ of the total area of the sky.
The concentration of both exoplanets and galactic radio sources in the galactic plane may alleviate this challenge somewhat, but it still appears formidable.

Observations of stellar radio emissions \citep{gudel2002} as an exoplanet transits in front of its parent star may offer a solution.
Here the stated requirement for the angular size of the distant astrophysical radio source to be no larger than the angular size over which the environment of the occulted planet is changing is clearly violated.
Nevertheless, imaging with suitable angular resolution can generate a small pixel size. 
The appropriate length-scale for Jupiter's inner magnetosphere 
is 2.5 $R_{J}$. At a distance of 10 light-years, that length-scale subtends an angle of 0.4 milliarcseconds. Very long baseline interferometry, such as the VLBA, offers such angular resolution at GHz frequencies.

Although not a direct result of the preceding work, one issue related to planetary ring systems should be emphasized.
If a nearby exoplanet with a Saturn-like ring system were to transit in front of its parent star, then very long baseline interferometry would have 
angular resolution comparable to the spatial extent of the ring system. Observed stellar radio emissions from the obscured region of the star would be appreciably attenuated.
For this to occur, the planet's ring plane must be tilted with respect to its orbital plane such that the line of sight from Earth to the planet is not parallel to the ring plane.


Occultation of a distant astrophysical radio source by a star, rather than a planet, may also be informative. The magnetic field and plasma in the extended stellar atmosphere will affect properties of the occulted radio signal just as in a planetary ionosphere and magnetosphere
\citep{tyler1977, woo1997, patzold2016}. 
Indeed, occultations of pulsars and other distant astrophysical radio sources by the Sun have been used to study the magnetic field and plasma density in the solar corona \citep{machin1952, bird1980, ingleby2007}.
Since known exoplanets number in the thousands, whereas the Gaia mission will observe $10^{9}$ stars, the likelihood of an occultation occurring within a reasonable period of time will be significantly greater for stars than for exoplanets. 
Moreover, high resolution imaging of radio emissions from one member of a binary star system as it is eclipsed by the other member would probe the extended stellar atmosphere of the nearer star.



\section{\label{sec:concs}Conclusions}

The polarization angle, frequency, and power of radio emissions from distant astrophysical radio sources, such as pulsars, active galactic nuclei, and masers, are affected by propagation through a planet's atmosphere, ionosphere, and magnetosphere.
Occultations of distant astrophysical radio sources by solar system planets, such as Jupiter, can be used to measure magnetic field strength, plasma density, and neutral density in planetary environments.
Based on the number of known distant astrophysical radio sources, such occultations are likely to occur often.
Occultations of distant astrophysical radio sources by exoplanets could probe the exoplanetary environments, but 
such occultations are unlikely to occur frequently.
An alternative type of occultation may be more promising for exoplanets: high resolution radio imaging of an exoplanet as it transits in front of its parent star. Stars in eclipsing binary systems could be studied similarly.

\indent{\bf{Acknowledgments\\}}

We acknowledge an anonymous reviewer.


\newpage



\begin{thebibliography}{}

\bibitem[{Acuna} et~al.(1983){Acuna}, {Behannon}, and {Connerney}]{acuna1983}
{Acuna}, M.~H., {Behannon}, K.~W., {Connerney}, J.~E.~P., 1983.
\newblock {Jupiter's magnetic field and magnetosphere}.
\newblock In: {Dessler}, A.~J. (Ed.), Physics of the jovian magnetosphere. pp.\
   1--50.

\bibitem[{Ahmad} and {Tyler}(1998){Ahmad} and {Tyler}]{ahmad1998}
{Ahmad}, B., {Tyler}, G.~L., 1998.
\newblock {The two-dimensional resolution kernel associated with retrieval of
  ionospheric and atmospheric refractivity profiles by Abelian inversion of
  radio occultation phase data}.
\newblock Radio Sci.~33, 129--142.

\bibitem[{Bagenal} and {Delamere}(2011){Bagenal} and {Delamere}]{bagenal2011}
{Bagenal}, F., {Delamere}, P.~A., 2011.
\newblock {Flow of mass and energy in the magnetospheres of Jupiter and
  Saturn}.
\newblock J. Geophys. Res.~116, A05209, 10.1029/2010JA016294.

\bibitem[{Behannon} et~al.(1981){Behannon}, {Burlaga}, and
  {Ness}]{behannon1981}
{Behannon}, K.~W., {Burlaga}, L.~F., {Ness}, N.~F., 1981.
\newblock {The Jovian magnetotail and its current sheet}.
\newblock J. Geophys. Res.~86, 8385--8401.

\bibitem[{Bird} et~al.(1980){Bird}, {Schruefer}, {Volland}, and
  {Sieber}]{bird1980}
{Bird}, M.~K., {Schruefer}, E., {Volland}, H., {Sieber}, W., 1980.
\newblock {Coronal Faraday rotation during solar occultation of PSR0525+21}.
\newblock Nature~283, 459--460.

\bibitem[{Bolton} et~al.(2004){Bolton}, {Thorne}, {Bourdarie}, {de Pater}, and
  {Mauk}]{bolton2004}
{Bolton}, S.~J., {Thorne}, R.~M., {Bourdarie}, S., {de Pater}, I., {Mauk}, B.,
  2004.
\newblock {Jupiter's inner radiation belts}.
\newblock In: {Bagenal}, F., {Dowling}, T.~E., {McKinnon}, W.~B. (Eds.),
  Jupiter.~The planet, satellites and magnetosphere. pp.\  671--688.

\bibitem[{Born} and {Wolf}(1959){Born} and {Wolf}]{born1959}
{Born}, M., {Wolf}, E., 1959.
\newblock {\em {Principles of optics}}.
\newblock Pergamon Press, London.

\bibitem[{Bouchez} et~al.(2003){Bouchez}, {Brown}, {Troy}, {Burruss}, {Dekany},
  and {West}]{bouchez2003}
{Bouchez}, A.~H., {Brown}, M.~E., {Troy}, M., {Burruss}, R.~S., {Dekany},
  R.~G., {West}, R.~A., 2003.
\newblock {Adaptive optics imaging of a stellar occultation by Titan}.
\newblock In: {Wizinowich}, P.~L., {Bonaccini}, D. (Eds.), Adaptive Optical
  System Technologies II, Volume 4839 of {\em \procspie}, pp.\  1045--1054.

\bibitem[Budden(1985)Budden]{budden1985}
Budden, K.~G., 1985.
\newblock {\em The propagation of radio waves}.
\newblock Cambridge University Press, Cambridge.

\bibitem[{Burke} and {Graham-Smith}(2010){Burke} and {Graham-Smith}]{burke2010}
{Burke}, B.~F., {Graham-Smith}, F., 2010.
\newblock {\em {An introduction to radio astronomy}}.
\newblock Cambridge University Press, New York.

\bibitem[{Carroll} and {Ostlie}(2007){Carroll} and {Ostlie}]{carroll2007}
{Carroll}, B.~W., {Ostlie}, D.~A., 2007.
\newblock {\em {An introduction to modern astrophysics, second edition}}.
\newblock Pearson, San Francisco.

\bibitem[{Counselman} and {Rankin}(1972){Counselman} and
  {Rankin}]{counselman1972}
{Counselman}, C.~C., III, {Rankin}, J.~M., 1972.
\newblock {Density of the solar corona from occultations of NP 0532}.
\newblock Astrophys. J.~175, 843.

\bibitem[{Counselman} and {Rankin}(1973){Counselman} and
  {Rankin}]{counselman1973}
{Counselman}, C.~C., III, {Rankin}, J.~M., 1973.
\newblock {Changes in the distribution of density and radio scattering in the
  solar corona in 1971}.
\newblock Astrophys. J.~185, 357--362.

\bibitem[{Croft}(1971){Croft}]{croft1971}
{Croft}, T.~A., 1971.
\newblock {Corotating regions in the solar wind, evident in number density
  measured by a radio-propagation technique}.
\newblock Radio Science~6, 55--63.

\bibitem[{de Pater} et~al.(2003){de Pater}, {Butler}, {Green}, {Strom},
  {Millan}, {Klein}, {Bird}, {Funke}, {Neidh{\"o}fer}, {Maddalena}, {Sault},
  {Kesteven}, {Smits}, and {Hunstead}]{depater2003}
{de Pater}, I., {Butler}, B.~J., {Green}, D.~A., {Strom}, R., {Millan}, R.,
  {Klein}, M.~J., {Bird}, M.~K., {Funke}, O., {Neidh{\"o}fer}, J., {Maddalena},
  R., {Sault}, R.~J., {Kesteven}, M., {Smits}, D.~P., {Hunstead}, R., 2003.
\newblock {Jupiter's radio spectrum from 74 MHz up to 8 GHz}.
\newblock Icarus~163, 434--448.

\bibitem[{Dias-Oliveira} et~al.(2015){Dias-Oliveira}, {Sicardy}, {Lellouch},
  {Vieira-Martins}, {Assafin}, {Camargo}, {Braga-Ribas}, {Gomes-J{\'u}nior},
  {Benedetti-Rossi}, {Colas}, {Decock}, {Doressoundiram}, {Dumas}, {Emilio},
  {Fabrega Polleri}, {Gil-Hutton}, {Gillon}, {Girard}, {Hau}, {Ivanov},
  {Jehin}, {Lecacheux}, {Leiva}, {Lopez-Sisterna}, {Mancini}, {Manfroid},
  {Maury}, {Meza}, {Morales}, {Nagy}, {Opitom}, {Ortiz}, {Pollock}, {Roques},
  {Snodgrass}, {Soulier}, {Thirouin}, {Vanzi}, {Widemann}, {Reichart},
  {LaCluyze}, {Haislip}, {Ivarsen}, {Dominik}, {J{\o}rgensen}, and
  {Skottfelt}]{diasoliveira2015}
{Dias-Oliveira}, A., {Sicardy}, B., {Lellouch}, E., {Vieira-Martins}, R.,
  {Assafin}, M., {Camargo}, J.~I.~B., {Braga-Ribas}, F., {Gomes-J{\'u}nior},
  A.~R., {Benedetti-Rossi}, G., {Colas}, F., {Decock}, A., {Doressoundiram},
  A., {Dumas}, C., {Emilio}, M., {Fabrega Polleri}, J., {Gil-Hutton}, R.,
  {Gillon}, M., {Girard}, J.~H., {Hau}, G.~K.~T., {Ivanov}, V.~D., {Jehin}, E.,
  {Lecacheux}, J., {Leiva}, R., {Lopez-Sisterna}, C., {Mancini}, L.,
  {Manfroid}, J., {Maury}, A., {Meza}, E., {Morales}, N., {Nagy}, L., {Opitom},
  C., {Ortiz}, J.~L., {Pollock}, J., {Roques}, F., {Snodgrass}, C., {Soulier},
  J.~F., {Thirouin}, A., {Vanzi}, L., {Widemann}, T., {Reichart}, D.~E.,
  {LaCluyze}, A.~P., {Haislip}, J.~B., {Ivarsen}, K.~M., {Dominik}, M.,
  {J{\o}rgensen}, U., {Skottfelt}, J., 2015.
\newblock {Pluto's atmosphere from stellar occultations in 2012 and 2013}.
\newblock Astrophys. J.~811, 53.

\bibitem[{Ebert} et~al.(2014){Ebert}, {Bagenal}, {McComas}, and
  {Fowler}]{ebert2014}
{Ebert}, R., {Bagenal}, F., {McComas}, D., {Fowler}, C., 2014.
\newblock {A survey of solar wind conditions at 5 AU: A tool for interpreting
  solar wind-magnetosphere interactions at Jupiter}.
\newblock Frontiers in Astronomy and Space Sciences~1, 1--13.

\bibitem[{Elliot} et~al.(1977){Elliot}, {Dunham}, and {Mink}]{elliot1977}
{Elliot}, J.~L., {Dunham}, E., {Mink}, D., 1977.
\newblock {The rings of Uranus}.
\newblock Nature~267, 328--330.

\bibitem[{Eshleman}(1973){Eshleman}]{eshleman1973}
{Eshleman}, V.~R., 1973.
\newblock {The radio occultation method for the study of planetary
  atmospheres}.
\newblock Planet. Space Sci.~21, 1521--1531.

\bibitem[{Eshleman} et~al.(1980){Eshleman}, {Steffes}, {Muhleman}, and
  {Nicholson}]{eshleman1980}
{Eshleman}, V.~R., {Steffes}, P.~G., {Muhleman}, D.~O., {Nicholson}, P.~D.,
  1980.
\newblock {Comment on absorbing regions in the atmosphere of Venus as measured
  by radio occultation}.
\newblock Icarus~44, 793--803.

\bibitem[{Eshleman} et~al.(1979){Eshleman}, {Tyler}, and
  {Freeman}]{eshleman1979}
{Eshleman}, V.~R., {Tyler}, G.~L., {Freeman}, W.~T., 1979.
\newblock {Deep radio occultations and ``evolute flashes'' --- Their
  characteristics and utility for planetary studies}.
\newblock Icarus~37, 612--626.

\bibitem[{Fjeldbo} and {Eshleman}(1965){Fjeldbo} and {Eshleman}]{fjeldbo1965a}
{Fjeldbo}, G., {Eshleman}, V.~R., 1965.
\newblock {The bistatic radar-occultation method for the study of planetary
  atmospheres}.
\newblock J. Geophys. Res.~70, 3217--3225.

\bibitem[{Fjeldbo} et~al.(1971){Fjeldbo}, {Kliore}, and
  {Eshleman}]{fjeldbo1971}
{Fjeldbo}, G., {Kliore}, A.~J., {Eshleman}, V.~R., 1971.
\newblock {The neutral atmosphere of Venus as studied with the Mariner V radio
  occultation experiments}.
\newblock Astron. J.~76, 123--140.

\bibitem[{Gray}(1999){Gray}]{gray1999}
{Gray}, M., 1999.
\newblock {Astrophysical masers}.
\newblock Philosophical Transactions of the Royal Society of London Series
  A~357, 3277--3298.

\bibitem[{G{\"u}del}(2002){G{\"u}del}]{gudel2002}
{G{\"u}del}, M., 2002.
\newblock {Stellar radio astronomy: Probing stellar atmospheres from protostars
  to giants}.
\newblock Ann. Rev. Astron. Astrophys.~40, 217--261.

\bibitem[Hinson et~al.(1999)Hinson, Simpson, Twicken, Tyler, and
  Flasar]{hinson1999}
Hinson, D.~P., Simpson, R.~A., Twicken, J.~D., Tyler, G.~L., Flasar, F.~M.,
  1999.
\newblock Initial results from radio occultation measurements with {M}ars
  {G}lobal {S}urveyor.
\newblock J. Geophys. Res.~104, 26997--27012.

\bibitem[{Howard} et~al.(1992){Howard}, {Eshleman}, {Hinson}, {Kliore},
  {Lindal}, {Woo}, {Bird}, {Volland}, {Edenhoffer}, and
  {P{\"a}tzold}]{howard1992}
{Howard}, H.~T., {Eshleman}, V.~R., {Hinson}, D.~P., {Kliore}, A.~J., {Lindal},
  G.~F., {Woo}, R., {Bird}, M.~K., {Volland}, H., {Edenhoffer}, P.,
  {P{\"a}tzold}, M., 1992.
\newblock {Galileo radio science investigations}.
\newblock Space Sci. Rev.~60, 565--590.

\bibitem[{Hubbard} et~al.(1975){Hubbard}, {Hunten}, and {Kliore}]{hubbard1975}
{Hubbard}, W.~B., {Hunten}, D.~M., {Kliore}, A., 1975.
\newblock {Effect of the Jovian oblateness on Pioneer 10/11 radio
  occultations}.
\newblock Geophys. Res. Lett.~2, 265--268.

\bibitem[{Ingleby} et~al.(2007){Ingleby}, {Spangler}, and
  {Whiting}]{ingleby2007}
{Ingleby}, L.~D., {Spangler}, S.~R., {Whiting}, C.~A., 2007.
\newblock {Probing the large-scale plasma structure of the solar corona with
  Faraday rotation measurements}.
\newblock Astrophys. J.~668, 520--532.

\bibitem[Jackson(1975)Jackson]{jackson1975}
Jackson, J.~D., 1975.
\newblock {\em Classical Electrodynamics, 2nd edition}.
\newblock Wiley, New York.

\bibitem[{Joy} et~al.(2002){Joy}, {Kivelson}, {Walker}, {Khurana}, {Russell},
  and {Ogino}]{joy2002}
{Joy}, S.~P., {Kivelson}, M.~G., {Walker}, R.~J., {Khurana}, K.~K., {Russell},
  C.~T., {Ogino}, T., 2002.
\newblock {Probabilistic models of the Jovian magnetopause and bow shock
  locations}.
\newblock J. Geophys. Res.~107, 1309, 10.1029/2001JA009146.

\bibitem[{Karayel} and {Hinson}(1997){Karayel} and {Hinson}]{karayel1997}
{Karayel}, E.~T., {Hinson}, D.~P., 1997.
\newblock {Sub-Fresnel-scale vertical resolution in atmospheric profiles from
  radio occultation}.
\newblock Radio Science~32, 411--424.

\bibitem[{Khurana} et~al.(2004){Khurana}, {Kivelson}, {Vasyliunas}, {Krupp},
  {Woch}, {Lagg}, {Mauk}, and {Kurth}]{khurana2004}
{Khurana}, K.~K., {Kivelson}, M.~G., {Vasyliunas}, V.~M., {Krupp}, N., {Woch},
  J., {Lagg}, A., {Mauk}, B.~H., {Kurth}, W.~S., 2004.
\newblock {The configuration of Jupiter's magnetosphere}.
\newblock In: {Bagenal}, F., {Dowling}, T.~E., {McKinnon}, W.~B. (Eds.),
  Jupiter.~The Planet, Satellites and Magnetosphere. pp.\  593--616.

\bibitem[{Khurana} and {Schwarzl}(2005){Khurana} and {Schwarzl}]{khurana2005}
{Khurana}, K.~K., {Schwarzl}, H.~K., 2005.
\newblock {Global structure of Jupiter's magnetospheric current sheet}.
\newblock J. Geophys. Res.~110, A07227, 10.1029/2004JA010757.

\bibitem[{Kivelson} and {Khurana}(2002){Kivelson} and {Khurana}]{kivelson2002}
{Kivelson}, M.~G., {Khurana}, K.~K., 2002.
\newblock {Properties of the magnetic field in the Jovian magnetotail}.
\newblock J. Geophys. Res.~107, 1196, 10.1029/2001JA000249.

\bibitem[{Kliore} et~al.(1975){Kliore}, {Fjeldbo}, {Seidel}, {Sesplaukis},
  {Sweetnam}, and {Woiceshyn}]{kliore1975b}
{Kliore}, A., {Fjeldbo}, G., {Seidel}, B.~L., {Sesplaukis}, T.~T., {Sweetnam},
  D.~W., {Woiceshyn}, P.~M., 1975.
\newblock {Atmosphere of Jupiter from the Pioneer 11 S-band occultation
  experiment --- Preliminary results}.
\newblock Science~188, 474--476.

\bibitem[{Kliore} et~al.(2004){Kliore}, {Anderson}, {Armstrong}, {Asmar},
  {Hamilton}, {Rappaport}, {Wahlquist}, {Ambrosini}, {Flasar}, {French},
  {Iess}, {Marouf}, and {Nagy}]{kliore2004}
{Kliore}, A.~J., {Anderson}, J.~D., {Armstrong}, J.~W., {Asmar}, S.~W.,
  {Hamilton}, C.~L., {Rappaport}, N.~J., {Wahlquist}, H.~D., {Ambrosini}, R.,
  {Flasar}, F.~M., {French}, R.~G., {Iess}, L., {Marouf}, E.~A., {Nagy}, A.~F.,
  2004.
\newblock {Cassini Radio Science}.
\newblock Space Sci. Rev.~115, 1--70.

\bibitem[{Kliore} et~al.(1976){Kliore}, {Woiceshyn}, and {Hubbard}]{kliore1976}
{Kliore}, A.~J., {Woiceshyn}, P.~M., {Hubbard}, W.~B., 1976.
\newblock {Temperature of the atmosphere of Jupiter from Pioneer 10/11 radio
  occultations}.
\newblock Geophys. Res. Lett.~3, 113--116.

\bibitem[{Krolik}(1999){Krolik}]{krolik1999}
{Krolik}, J.~H., 1999.
\newblock {\em {Active galactic nuclei: From the central black hole to the
  galactic environment}}.
\newblock Princeton University Press, Princeton.

\bibitem[{Lepping} et~al.(1983){Lepping}, {Desch}, {Sittler}, {Behannon},
  {Klein}, {Sullivan}, and {Kurth}]{lepping1983}
{Lepping}, R.~P., {Desch}, M.~D., {Sittler}, E.~C., Jr., {Behannon}, K.~W.,
  {Klein}, L.~W., {Sullivan}, J.~D., {Kurth}, W.~S., 1983.
\newblock {Structure and other properties of Jupiter's distant magnetotail}.
\newblock J. Geophys. Res.~88, 8801--8815.

\bibitem[{Lindal} et~al.(1983){Lindal}, {Wood}, {Hotz}, {Sweetnam}, {Eshleman},
  and {Tyler}]{lindal1983}
{Lindal}, G.~F., {Wood}, G.~E., {Hotz}, H.~B., {Sweetnam}, D.~N., {Eshleman},
  V.~R., {Tyler}, G.~L., 1983.
\newblock {The atmosphere of Titan - an analysis of the Voyager 1 radio
  occultation measurements}.
\newblock Icarus~53, 348--363.

\bibitem[{Lo}(2005){Lo}]{lo2005}
{Lo}, K.~Y., 2005.
\newblock {Mega-masers and galaxies}.
\newblock Ann. Rev. Astron. Astrophys.~43, 625--676.

\bibitem[{Machin} and {Smith}(1952){Machin} and {Smith}]{machin1952}
{Machin}, K.~E., {Smith}, F.~G., 1952.
\newblock {Occultation of a radio star by the solar corona}.
\newblock Nature~170, 319--320.

\bibitem[{Manchester} et~al.(2005){Manchester}, {Hobbs}, {Teoh}, and
  {Hobbs}]{manchester2005}
{Manchester}, R.~N., {Hobbs}, G.~B., {Teoh}, A., {Hobbs}, M., 2005.
\newblock {The Australia Telescope National Facility Pulsar Catalogue}.
\newblock Astron. J.~129, 1993--2006.

\bibitem[{Nicolaou} et~al.(2015){Nicolaou}, {McComas}, {Bagenal}, {Elliott},
  and {Wilson}]{nicolaou2015}
{Nicolaou}, G., {McComas}, D.~J., {Bagenal}, F., {Elliott}, H.~A., {Wilson},
  R.~J., 2015.
\newblock {Plasma properties in the deep jovian magnetotail}.
\newblock Planet. Space Sci.~119, 222--232.

\bibitem[{Ord} et~al.(2007){Ord}, {Johnston}, and {Sarkissian}]{ord2007}
{Ord}, S.~M., {Johnston}, S., {Sarkissian}, J., 2007.
\newblock {The magnetic field of the solar corona from pulsar observations}.
\newblock Solar Phys.~245, 109--120.

\bibitem[{P{\"a}tzold} et~al.(2016){P{\"a}tzold}, {H{\"a}usler}, {Tyler},
  {Andert}, {Asmar}, {Bird}, {Dehant}, {Hinson}, {Rosenblatt}, {Simpson},
  {Tellmann}, {Withers}, {Beuthe}, {Efimov}, {Hahn}, {Kahan}, {Le Maistre},
  {Oschlisniok}, {Peter}, and {Remus}]{patzold2016}
{P{\"a}tzold}, M., {H{\"a}usler}, B., {Tyler}, G.~L., {Andert}, T., {Asmar},
  S.~W., {Bird}, M.~K., {Dehant}, V., {Hinson}, D.~P., {Rosenblatt}, P.,
  {Simpson}, R.~A., {Tellmann}, S., {Withers}, P., {Beuthe}, M., {Efimov},
  A.~I., {Hahn}, M., {Kahan}, D., {Le Maistre}, S., {Oschlisniok}, J., {Peter},
  K., {Remus}, S., 2016.
\newblock {Mars Express 10 years at Mars: Observations by the Mars Express
  Radio Science Experiment (MaRS)}.
\newblock Planet. Space Sci.~127, 44--90.

\bibitem[{Qu{\'e}merais} et~al.(2006){Qu{\'e}merais}, {Bertaux}, {Korablev},
  {Dimarellis}, {Cot}, {Sandel}, and {Fussen}]{quemerais2006}
{Qu{\'e}merais}, E., {Bertaux}, J.-L., {Korablev}, O., {Dimarellis}, E., {Cot},
  C., {Sandel}, B.~R., {Fussen}, D., 2006.
\newblock {Stellar occultations observed by SPICAM on Mars Express}.
\newblock J. Geophys. Res.~111, E09S04, 10.1029/2005JE002604.

\bibitem[{Seiff} et~al.(1998){Seiff}, {Kirk}, {Knight}, {Young}, {Mihalov},
  {Young}, {Milos}, {Schubert}, {Blanchard}, and {Atkinson}]{seiff1998}
{Seiff}, A., {Kirk}, D.~B., {Knight}, T.~C.~D., {Young}, R.~E., {Mihalov},
  J.~D., {Young}, L.~A., {Milos}, F.~S., {Schubert}, G., {Blanchard}, R.~C.,
  {Atkinson}, D., 1998.
\newblock {Thermal structure of Jupiter's atmosphere near the edge of a
  5-{$\mu$}m hot spot in the north equatorial belt}.
\newblock J. Geophys. Res.~103, 22857--22890.

\bibitem[{Sicardy} et~al.(2006){Sicardy}, {Colas}, {Widemann}, {Bellucci},
  {Beisker}, {Kretlow}, {Ferri}, {Lacour}, {Lecacheux}, {Lellouch}, {Pau},
  {Renner}, {Roques}, {Fienga}, {Etienne}, {Martinez}, {Glass}, {Baba},
  {Nagayama}, {Nagata}, {Itting-Enke}, {Bath}, {Bode}, {Bode}, {L{\"u}demann},
  {L{\"u}demann}, {Neubauer}, {Tegtmeier}, {Tegtmeier}, {Thom{\'e}}, {Hund},
  {deWitt}, {Fraser}, {Jansen}, {Jones}, {Schoenau}, {Turk}, {Meintjies},
  {Hernandez}, {Fiel}, {Frappa}, {Peyrot}, {Teng}, {Vignand}, {Hesler},
  {Payet}, {Howell}, {Kidger}, {Ortiz}, {Naranjo}, {Rosenzweig}, and
  {Rapaport}]{sicardy2006}
{Sicardy}, B., {Colas}, F., {Widemann}, T., {Bellucci}, A., {Beisker}, W.,
  {Kretlow}, M., {Ferri}, F., {Lacour}, S., {Lecacheux}, J., {Lellouch}, E.,
  {Pau}, S., {Renner}, S., {Roques}, F., {Fienga}, A., {Etienne}, C.,
  {Martinez}, C., {Glass}, I.~S., {Baba}, D., {Nagayama}, T., {Nagata}, T.,
  {Itting-Enke}, S., {Bath}, K.-L., {Bode}, H.-J., {Bode}, F., {L{\"u}demann},
  H., {L{\"u}demann}, J., {Neubauer}, D., {Tegtmeier}, A., {Tegtmeier}, C.,
  {Thom{\'e}}, B., {Hund}, F., {deWitt}, C., {Fraser}, B., {Jansen}, A.,
  {Jones}, T., {Schoenau}, P., {Turk}, C., {Meintjies}, P., {Hernandez}, M.,
  {Fiel}, D., {Frappa}, E., {Peyrot}, A., {Teng}, J.~P., {Vignand}, M.,
  {Hesler}, G., {Payet}, T., {Howell}, R.~R., {Kidger}, M., {Ortiz}, J.~L.,
  {Naranjo}, O., {Rosenzweig}, P., {Rapaport}, M., 2006.
\newblock {The two Titan stellar occultations of 14 November 2003}.
\newblock J. Geophys. Res.~111, E11S91, 10.1029/2005JE002624.

\bibitem[{Sicardy} et~al.(2016){Sicardy}, {Talbot}, {Meza}, {Camargo},
  {Desmars}, {Gault}, {Herald}, {Kerr}, {Pavlov}, {Braga-Ribas}, {Assafin},
  {Benedetti-Rossi}, {Dias-Oliveira}, {Gomes-J{\'u}nior}, {Vieira-Martins},
  {B{\'e}rard}, {Kervella}, {Lecacheux}, {Lellouch}, {Beisker}, {Dunham},
  {Jel{\'{\i}}nek}, {Duffard}, {Ortiz}, {Castro-Tirado}, {Cunniffe}, {Querel},
  {Yock}, {Cole}, {Giles}, {Hill}, {Beaulieu}, {Harnisch}, {Jansen}, {Pennell},
  {Todd}, {Allen}, {Graham}, {Loader}, {McKay}, {Milner}, {Parker}, {Barry},
  {Bradshaw}, {Broughton}, {Davis}, {Devillepoix}, {Drummond}, {Field},
  {Forbes}, {Giles}, {Glassey}, {Groom}, {Hooper}, {Horvat}, {Hudson},
  {Idaczyk}, {Jenke}, {Lade}, {Newman}, {Nosworthy}, {Purcell}, {Skilton},
  {Streamer}, {Unwin}, {Watanabe}, {White}, and {Watson}]{sicardy2016}
{Sicardy}, B., {Talbot}, J., {Meza}, E., {Camargo}, J.~I.~B., {Desmars}, J.,
  {Gault}, D., {Herald}, D., {Kerr}, S., {Pavlov}, H., {Braga-Ribas}, F.,
  {Assafin}, M., {Benedetti-Rossi}, G., {Dias-Oliveira}, A.,
  {Gomes-J{\'u}nior}, A.~R., {Vieira-Martins}, R., {B{\'e}rard}, D.,
  {Kervella}, P., {Lecacheux}, J., {Lellouch}, E., {Beisker}, W., {Dunham}, D.,
  {Jel{\'{\i}}nek}, M., {Duffard}, R., {Ortiz}, J.~L., {Castro-Tirado}, A.~J.,
  {Cunniffe}, R., {Querel}, R., {Yock}, P.~C., {Cole}, A.~A., {Giles}, A.~B.,
  {Hill}, K.~M., {Beaulieu}, J.~P., {Harnisch}, M., {Jansen}, R., {Pennell},
  A., {Todd}, S., {Allen}, W.~H., {Graham}, P.~B., {Loader}, B., {McKay}, G.,
  {Milner}, J., {Parker}, S., {Barry}, M.~A., {Bradshaw}, J., {Broughton}, J.,
  {Davis}, L., {Devillepoix}, H., {Drummond}, J., {Field}, L., {Forbes}, M.,
  {Giles}, D., {Glassey}, R., {Groom}, R., {Hooper}, D., {Horvat}, R.,
  {Hudson}, G., {Idaczyk}, R., {Jenke}, D., {Lade}, B., {Newman}, J.,
  {Nosworthy}, P., {Purcell}, P., {Skilton}, P.~F., {Streamer}, M., {Unwin},
  M., {Watanabe}, H., {White}, G.~L., {Watson}, D., 2016.
\newblock {Pluto's atmosphere from the 2015 June 29 ground-based stellar
  occultation at the time of the New Horizons flyby}.
\newblock Astrophys. J. Lett.~819, L38.

\bibitem[{Tyler}(1987){Tyler}]{tyler1987}
{Tyler}, G.~L., 1987.
\newblock {Radio propagation experiments in the outer solar system with
  Voyager}.
\newblock IEEE Proceedings~75, 1404--1431.

\bibitem[{Tyler} et~al.(1977){Tyler}, {Brenkle}, {Komarek}, and
  {Zygielbaum}]{tyler1977}
{Tyler}, G.~L., {Brenkle}, J.~P., {Komarek}, T.~A., {Zygielbaum}, A.~I., 1977.
\newblock {The Viking solar corona experiment}.
\newblock J. Geophys. Res.~82, 4335--4340.

\bibitem[{Weisberg} et~al.(1976){Weisberg}, {Rankin}, {Payne}, and
  {Counselman}]{weisberg1976}
{Weisberg}, J.~M., {Rankin}, J.~M., {Payne}, R.~R., {Counselman}, C.~C., III,
  1976.
\newblock {Further changes in the distribution of density and radio scattering
  in the solar corona in 1973}.
\newblock Astrophys. J.~209, 252--258.

\bibitem[{Withers}(2010){Withers}]{withers2010}
{Withers}, P., 2010.
\newblock {Prediction of uncertainties in atmospheric properties measured by
  radio occultation experiments}.
\newblock Adv. Space Res.~46, 58--73.

\bibitem[{Withers}(2011){Withers}]{withers2011}
{Withers}, P., 2011.
\newblock {Attenuation of radio signals by the ionosphere of Mars: Theoretical
  development and application to MARSIS observations}.
\newblock Radio Sci.~46, RS2004, 10.1029/2010RS004450.

\bibitem[{Withers} et~al.(2014){Withers}, {Moore}, {Cahoy}, and
  {Beerer}]{withers2014}
{Withers}, P., {Moore}, L., {Cahoy}, K., {Beerer}, I., 2014.
\newblock {How to process radio occultation data: 1. From time series of
  frequency residuals to vertical profiles of atmospheric and ionospheric
  properties}.
\newblock Planet. Space Sci.~101, 77--88.

\bibitem[{Woo}(1997){Woo}]{woo1997}
{Woo}, R., 1997.
\newblock {Morphology of the solar corona from radio occultation measurements:
  Implications for Solar Probe}.
\newblock In: {Habbal}, S.~R. (Ed.), Robotic Exploration Close to the Sun:
  Scientific Basis, Volume 385 of {\em American Institute of Physics Conference
  Series}, pp.\  85--96.

\bibitem[{Yakovlev}(2002){Yakovlev}]{yakovlev2002}
{Yakovlev}, O.~I., 2002.
\newblock {\em {Space radio science}}.
\newblock Taylor and Francis, New York.

\bibitem[{Yelle} and {Miller}(2004){Yelle} and {Miller}]{yelle2004}
{Yelle}, R.~V., {Miller}, S., 2004.
\newblock {Jupiter's thermosphere and ionosphere}.
\newblock In: {Bagenal}, F., {Dowling}, T.~E., {McKinnon}, W.~B. (Eds.),
  Jupiter.~The Planet, Satellites and Magnetosphere. pp.\  185--218.

\end{thebibliography}

\newpage

\clearpage

\clearpage

\clearpage

\clearpage

\clearpage

\end{document}